\documentstyle[11pt,epsfig]{article}
\def\be{\begin{equation}}
\def\ee{\end{equation}}
\def\bc{\begin{center}}
\def\ec{\end{center}}
\def\bea{\begin{eqnarray}}
\def\eea{\end{eqnarray}}

\def\nn{\nonumber}

\def\ov{\overline}
\def\as{\alpha_s}
\def\at{\alpha_t}
\def\sq2{\sqrt{2}}
\def\ths{\bar{\theta}_{\tilde{t}}}
\def\thz{\theta_{\tilde{t}}}
\def\mix{\widetilde{X}}
\def\mixt{X}
\def\c2t2{c_{2\bar{\theta}}^{\,2}}
\def\csenza{c_{2 \theta}^{\,2}}
\def\s2t{s_{2\theta}}

\def\mgl{m_{\tilde{g}}}
\def\msqu{m_{\tilde{t}_1}^2}
\def\msqd{m_{\tilde{t}_2}^2}
\def\msqc{m_{\tilde{t}}^2}
\def\mtu{m_{\tilde{t}_1}}
\def\mtd{m_{\tilde{t}_2}}
\def\cpp{c_{\varphi\tilde{\varphi}}}
\def\v2lp{V^{\rm 2loop}}
\def\veff{V_{\rm eff}}
\def\mt{m_t}
\def\msquq{m_{\tilde{t}_1}^4}
\def\msqdq{m_{\tilde{t}_2}^4}
\def\diff{\msqu-\msqd}
\def\mylg{\ln\frac{\mgl^2}{Q^2}}
\def\mylt{\ln\frac{m_t^2}{Q^2}}
\def\myltu{\ln\frac{\msqu}{Q^2}}
\def\myltd{\ln\frac{\msqd}{Q^2}}

\def\myltq{\ln^2\frac{m_t^2}{Q^2}}
\def\myltuq{\ln^2\frac{\msqu}{Q^2}}
\def\myltdq{\ln^2\frac{\msqd}{Q^2}}
\def\den{\Delta}

\newenvironment{appendletterA}
 {
  \setcounter{section}{0}
  \setcounter{equation}{0}
  
 }{
 }
\newenvironment{appendletterB}
 {
  \setcounter{equation}{0}
  
 }{
 }
\newenvironment{appendletterC}
 {
  \setcounter{equation}{0}
  
 }{
 }
\begin{document}
\begin{titlepage}
\vspace*{-1cm}
\phantom{hep-ph/0105096} 
\hfill{DFPD-01/TH/16}
\\
\phantom{hep-ph/0105096} 
\hfill{ROME1-1312-01}
\vskip 2.0cm
\begin{center}
{\Large\bf On the neutral Higgs boson masses in \\
           the MSSM for arbitrary stop mixing}
\end{center}
\vskip 1.5  cm
\begin{center}
{\large Giuseppe Degrassi}~\footnote{e-mail address: degrassi@pd.infn.it},
\hspace*{0.5cm}
{\large Pietro Slavich}~\footnote{e-mail address: slavich@pd.infn.it}
\\
\vskip .1cm
Dipartimento di Fisica `G.~Galilei', Universit\`a di Padova and
\\ 
INFN, Sezione di Padova, Via Marzolo~8, I-35131 Padua, Italy
\\
\vskip .2cm
and
\\
\vskip .2cm
{\large Fabio Zwirner}~\footnote{e-mail address: fabio.zwirner@roma1.infn.it}
\\
\vskip .1cm
Dipartimento di Fisica, Universit\`a di Roma `La Sapienza' and
\\
INFN, Sezione di Roma, P.le Aldo Moro~2, I-00185 Rome, Italy
\end{center}
\vskip 2.5cm
\begin{abstract}
\noindent
We compute the ${\cal O}(\at \as)$ two--loop corrections to 
the neutral Higgs boson masses in the Minimal Supersymmetric extension 
of the Standard Model. An appropriate use of the effective potential 
allows us to obtain simple analytical formulae, valid for arbitrary 
values of $m_A$ and of the mass parameters in the stop sector. We 
elucidate some subtleties of the effective potential calculation, 
and find full agreement with the numerical output of the existing
diagrammatic calculation. We discuss in detail the limit of heavy 
gluino.
\end{abstract}
\end{titlepage}
\setcounter{footnote}{0}
\vskip2truecm


\section{Introduction}
\label{sect:intro}
There is a crucial prediction of the Minimal Supersymmetric 
extension of the Standard Model, or MSSM (for reviews and 
references, see e.g. ref.~\cite{mssm}), subject to decisive
tests at present and future colliders.
It is the existence of a light CP--even neutral Higgs boson,
$h$, accompanied by other states $(H,A,H^\pm)$, whose masses
are strongly correlated but can vary over a wide range near the 
weak scale. An important step in the understanding of the MSSM
Higgs sector was the realization that the classical bound $m_h
< m_Z$, and, more generally, the classical relations among the 
gauge and Higgs boson masses, are violated by large radiative 
corrections, dominated by top and stop loops \cite{erz1,oyy1,hh1}. 
After that, extensive efforts have been devoted to progressive 
refinements of the theoretical predictions for the Higgs boson 
masses and couplings, as functions of the relevant MSSM parameters. 
These activities have been performed in several directions, with 
special emphasis on the prediction for $m_h$: inclusion of stop 
mixing effects \cite{oyy2,erz2}; resummation of large logarithms 
using appropriate one-- and two--loop renormalization group equations (RGE) 
\cite{oyy2,bfc,rge,espnav}; 
complete one--loop diagrammatic calculations including momentum-dependent 
effects \cite{diag1,bagger}; calculations of the most important 
two--loop contributions \cite{hh3,hhw2l,combi,ez}. Other studies have 
been oriented towards a meaningful combination of the above results 
\cite{ez,reconciling}, and towards the implementation of the latter
in computer codes \cite{codes,FeyH}, to be used in turn for experimental 
analyses \cite{exp}.

In the present paper we address once more the computation of the 
neutral Higgs boson masses, whose present state--of--the--art can 
be summarized as follows. There is a  diagrammatic two--loop 
computation, including ${\cal O} (\at \as)$ effects, 
performed for arbitrary $m_A$ and arbitrary values of the parameters
in the stop mass matrix, in the zero-momentum limit \cite{hhw2l}. 
While, for small stop mixing and universal soft stop masses, 
sufficiently simple and accurate analytical formulae have been 
obtained \cite{combi}, in the general case the complete formulae 
are rather lenghty, which may be a problem for their practical 
implementation in computer codes. The results of ref.~\cite{hhw2l} 
can be improved by including the logarithmic ${\cal O} (\at^2)$ 
corrections, as extracted by solving perturbatively the appropriate 
RGE \cite{rge,espnav}.  There is also a computation \cite{ez} of
both ${\cal O} (\at \as)$ and ${\cal O} (\at^2)$ two--loop 
corrections to $m_h$, based on the effective potential approach.
This computation, however, is applicable only for $m_A \gg m_Z$. 
Moreover, the full results of \cite{ez} for $m_h$ are 
available only in numerical form, and accurate and simple analytical 
formulae were provided under the additional assumptions of small 
stop mixing and universal soft stop masses. 

In view of the situation described above, there is still room 
for a number of useful improvements that could be achieved without
excessive effort. As a first step, one should aim at simple analytical 
formulae for the two--loop corrected mass matrix of the neutral 
CP--even Higgs sector, still in the zero--momentum limit and including 
only  ${\cal O} (\at \as)$ corrections, but for arbitrary 
values of $m_A$ and of the parameters of the stop mass matrix. One
could then proceed with the inclusion of the ${\cal O} (\at^2)$ 
corrections, and of the corrections coming from the momentum--dependent
part of the two--loop Higgs propagators, into the above framework.
Finally, one could address the resummation of the large logarithms of 
$(\mtu/\mtd$), the ratio of the two stop 
mass eigenvalues, by means of suitable RGE, defined in an appropriate 
effective theory. This has been done \cite{espnav} in the case of small 
stop mixing, but is considerably more complicated in the case where
a large splitting between $\mtu$ and $\mtd$ is induced by 
a large mixing term in the stop mass matrix. 

In this paper we accomplish the first step of the above program,
leaving the remaining steps for future work. The paper is organized 
as follows. After this introduction, section~2 recalls the general
features of the calculation of the MSSM neutral Higgs boson masses 
in the effective potential approach. Section~3 describes the main 
features of our two--loop calculation of the ${\cal O} (\at \as)$  
contributions, and presents its results, in a form that allows
to assign the input parameters either in the $\ov{\rm DR}$ scheme
or in some on--shell scheme. In the concluding section we compare 
our results with the existing literature, and discuss in detail the 
heavy--gluino limit. This limit requires some care, especially
when the input parameters are assigned in the $\ov{\rm DR}$ scheme, 
as often done in models that predict the soft supersymmetry--breaking 
masses.

Technical details are confined to three appendices. Appendix~A gives 
the analytical expressions for the two--loop contributions to the 
neutral Higgs mass matrices that are controlled by the gluino mass. 
Appendix~B gives the explicit formulae that are needed for the 
transition from the $\ov{\rm DR}$ scheme to our implementation of 
the on--shell scheme. Appendix C gives the relation between $m_3^2$ 
and $m_A^2$, which may be useful for discussing models that predict 
the values of the soft supersymmetry--breaking masses, and in
particular of $m_3^2$, at some cut--off scale for the MSSM.
\section{Higgs masses in the effective potential approach}
\label{sect:higgsmass}
The MSSM neutral Higgs boson masses can be identified with the 
zeros of the corresponding two-point functions, which depend on
the external momentum and have in general the form of a matrix. 
In the limit of vanishing external momentum, these masses can be 
formally obtained by the following method: compute the effective 
potential $V_{\rm eff}$, retaining its complete dependence on the 
neutral Higgs fields, $H_1^0$ and $H_2^0$; minimize $V_{\rm eff}$ 
to find the vacuum expectation values $\langle H^0_1 \rangle \equiv
v_1/\sqrt{2}$ and $\langle H^0_2 \rangle \equiv v_2 / \sqrt{2}$; 
expand $V_{\rm eff}$ around its minumum up to quadratic fluctuations, 
and diagonalize the resulting mass matrix. Of course, at each step 
we can carry out explicitly only the calculations that contribute 
to the final results, at the desired level of approximation. Before 
moving to our specific two--loop computation, we give now some 
formulae that illustrate the formal results of the general 
procedure, and are valid at every order in perturbation theory.

Putting all the other fields to zero, and keeping only the dependence 
on the neutral Higgs fields, the tree--level Higgs potential of the 
MSSM reads:
\be
\label{Vtree}
V_0  =  
m_1^2 \, \left| H_1^0 \right|^2 
+ m_2^2 \, \left| H_2^0 \right|^2 
+ m_3^2 \, \left( H_1^0 H_2^0 + {\rm h.c.} \right)
+ {g^2 +g^{\prime\,2} \over 8} \left(
|H_1^0|^2 - |H_2^0|^2 \right)^2 \, ,
\ee
where: $m_1^2 = m_{H_1}^2 + |\mu|^2$, $m_2^2 = m_{H_2}^2 + |\mu|^2$; 
$\mu$ is the Higgs mass term in the superpotential; $m_{H_1}^2$, 
$m_{H_2}^2$ and 
$m_3^2$ are soft supersymmetry--breaking masses; $g$ and $g'$ are the 
$SU(2)_L$ and $U(1)_Y$ gauge couplings, respectively.
It is not restrictive to choose $m_3^2$ real and negative, so that 
$v_1$ and $v_2$ are real and positive, and the neutral Higgs fields 
can be decomposed into their vacuum expectation values plus their 
CP--even and CP--odd fluctuations as follows:
\be 
\label{expansion}
H_1^0 \equiv {v_1 + S_1 + i \, P_1 \over \sq2} \, ,
\;\;\;\;\;\;\;
H_2^0 \equiv {v_2 + S_2 + i \, P_2 \over \sq2} \, .
\ee

In the effective potential approach, the mass matrices for the neutral 
CP--odd and CP--even Higgs bosons can be approximated, at every order 
in perturbation theory, by:
\be
\label{defmat}
\left({\cal M}^2_P\right)_{ij} = 
\left. \frac{\partial^2  V_{\rm eff}}{\partial P_i \partial P_j}
\right|_{\rm min} \, , 
\hspace{1cm}
\left({\cal M}^2_S\right)_{ij} = 
\left. \frac{\partial^2  V_{\rm eff}}{\partial S_i \partial S_j}
\right|_{\rm min} \, ,
\hspace{1cm}
(i,j=1,2) \, ,
\ee
where $V_{\rm eff} = V_0 + V$ is the loop--corrected Higgs potential
in the $\ov{\rm DR}$ scheme, and $\langle S_1 \rangle = \langle P_1 
\rangle = \langle S_2 \rangle = \langle P_2 \rangle = 0$. Using the 
explicit expression of the tree--level potential, eq.~(\ref{Vtree}), 
$v_1$ and $v_2$ are determined by minimizing the effective potential:
\be
\label{minima1}
{1 \over v_1}
\left.\frac{\,\partial V_{\rm eff}}{\partial S_1} \right|_{\rm min} 
\; = \; 
m_1^2 + m_3^2 \,\frac{v_2}{v_1} + {(g^2 + g'^2) \over 4} (v_1^2 - v_2^2) 
+ 
\left.\frac{1}{v_1} \frac{\partial V}{\partial S_1}\right|_{\rm min} 
\; = \; 0 \, ,
\ee 
\be
\label{minima2}
{1 \over v_2}
\left.\frac{\,\partial V_{\rm eff}}{\partial S_2} \right|_{\rm min} 
\; = \; 
m_2^2 + m_3^2 \,\frac{v_1}{v_2} + {(g^2 + g'^2) \over 4} (v_2^2 - v_1^2) 
+ 
\left.\frac{1}{v_2} \frac{\partial V}{\partial S_2}\right|_{\rm min} 
\; = \; 0 \, .
\ee
Combining eqs.~(\ref{Vtree})--(\ref{minima2}), the CP--odd and 
CP--even Higgs mass matrices become $(i,j=1,2)$:
\bea
\label{mpij}
\left({\cal M}^2_P\right)_{ij} & = & 
- m_3^2 \, \frac{v_1 v_2}{v_i v_j}
-\left. \frac{\delta_{ij}}{v_i} 
\frac{\partial V}{\partial S_i}\right|_{\rm min}
+ \left.\frac{\partial^2 V}{\partial P_i \partial P_j}\right|_{\rm min}
\, , \\ & & \nn \\
\label{msij}
\left({\cal M}^2_S\right)_{ij} & = &
(-1)^{i+j}\left[- m_3^2 \, \frac{v_1 v_2}{v_i v_j} 
+ {(g^2 + g'^2) \over 2} v_i v_j \right] - \frac{\delta_{ij}}{v_i} 
\left.\frac{\partial V}{\partial S_i}\right|_{\rm min}
+ \left.\frac{\partial^2 V}{\partial S_i \partial S_j}\right|_{\rm min}
\,  .
\eea
Combining further eqs.~(\ref{mpij}) and (\ref{msij}), we can write the 
CP--even Higgs mass matrix as follows:
\be
\label{msij2}
\left({\cal M}^2_S\right)_{ij}  = 
(-1)^{i+j} \left[ \left({\cal M}^2_P\right)_{ij} 
-
\left.\frac{\partial^2 V}{\partial P_i \partial P_j}\right|_{\rm min}
+
{(g^2 + g'^2) \over 2} v_i v_j  \; \right] \;
+\left.\frac{\partial^2 V}{\partial S_i \partial S_j}\right|_{\rm min}
\, ,
\ee
where the CP-odd mass matrix can be expressed, in terms of the 
loop--corrected CP-odd Higgs mass $m_A$ and of $\tan \beta = v_2
/ v_1$, as:
\be
\label{mpma}
{\cal M}^2_P = 
\left(\begin{array}{cc}
\sin^2\beta & \sin\beta\,\cos\beta\\
\sin\beta\,\cos\beta & \cos^2\beta
\end{array}\right)\,m_A^2 \;.
\ee


\section{${\cal O}(\at \as)$ two-loop corrections to the 
neutral Higgs boson masses}

\begin{figure}
\begin{center}
\epsfig{figure=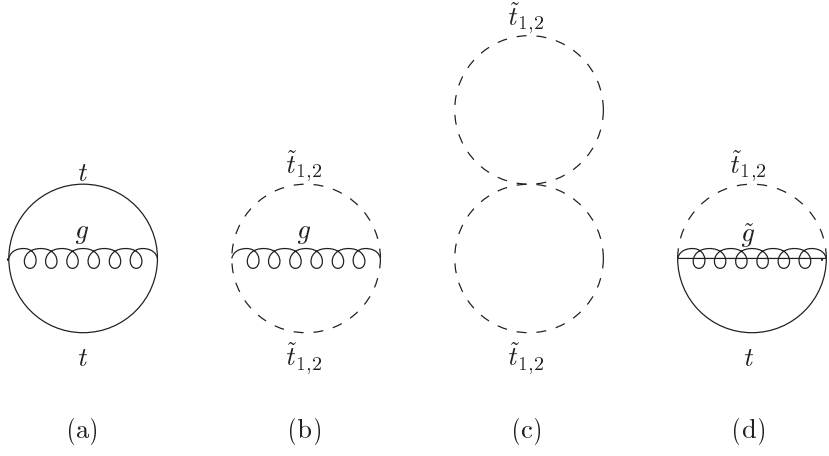,width=11cm}
\end{center}
\caption{Diagrams that contribute to the two--loop effective potential
and affect the ${\cal O}(\at \as)$ calculation of the neutral Higgs 
boson masses.}
\label{figdiag}
\end{figure}

The formulae derived in the previous section have general validity,
and were employed long ago for the one--loop computation \cite{erz1,
oyy1,oyy2,erz2}. We will now follow the same strategy for the calculation
of the ${\cal O}(\at \as)$ two--loop corrections to the entries of
the neutral CP--even Higgs boson mass matrix. The relevant Feynman 
diagrams involve top, stop, gluons and gluinos on the internal 
lines, and are shown in figure~\ref{figdiag}. Since $\veff$ must be 
considered in a generic Higgs background, it is important to elucidate 
the dependence of the propagators and vertices on the Higgs fields.

In a generic Higgs background, the MSSM Lagrangian contains the 
following bilinear terms in the top fields:
\be
{\cal L}_{2t} =
(\ov{t_L^{\, \prime}} \;\; \ov{t_R^{\, \prime}}) \;
\left(
\begin{array}{cc}
i \not \! \partial & - \mixt^*\\ -\mixt & i \not \! \partial 
\end{array}
\right)
\;
\left(
\begin{array}{c}
t_L^{\, \prime} \\ t_R^{\, \prime}
\end{array}
\right) \, ,
\ee
where $t_L^{\, \prime}$ and $t_R^{\, \prime}$ are four--component 
fermions of definite chirality, and the field--dependent mixing 
term $\mixt$ is:
\be
\label{mixt}
\mixt \; = \; h_t\, H^0_2 \; \equiv \; |\mixt|  \, e^{i\,\varphi}
\,,
\hspace{1.5cm}
(0 \leq \varphi < 2 \pi) \, .
\ee
It is not restrictive to assume that $h_t$ is real and positive,
then $\langle |\mixt| \rangle = h_t v_2 / \sqrt{2}$ and $\langle 
\varphi \rangle = 0$. Analogously, the terms quadratic in the top 
squarks are:
\be
{\cal L}_{2\tilde{t}} = -
(\tilde{t}^{\,\prime\,*}_L \;\; \tilde{t}^{\,\prime\,*}_R) \;
\left(
\begin{array}{cc}
\Box + m_L^2 & \mix^*\\
\mix & \Box + m_R^2 
\end{array}
\right)\;
\left(
\begin{array}{c}
\tilde{t}_L^{\, \prime} \\ \tilde{t}_R^{\, \prime}
\end{array}
\right)
\, ,
\ee
and the field--dependent entries of the stop mass matrix are,
neglecting D--term contributions that vanish for $g=g'=0$:
\be
m_L^2 = m_Q^2 + h_t^2 \, |H_2^0|^2 \, ,
\hspace{1cm}
m_R^2 = m_U^2 + h_t^2 \, |H_2^0|^2 \, ,
\ee
\be
\label{mix}
\mix  \; \equiv |\mix|\, e^{i\, \widetilde{\varphi}} \;  =  \;  
h_t\,\left( A_t H_2^0 + \mu H_1^{0\,*}\right) \, ,
\hspace{1.cm}
(0 \leq \widetilde{\varphi} < 2 \pi) \, ,
\ee
where $m_Q^2$, $m_U^2$ and $A_t$ are the soft supersymmetry--breaking 
mass parameters of the stop sector. We assume here $\mu$ and $A_t$ to
be real, so that $\langle |\mix| \rangle = (h_t v_2/\sqrt{2}) \,
|A_t + \mu \cot\beta|$ and $\langle e^{i\, \widetilde{\varphi}} \rangle 
= sign \, (A_t + \mu \cot\beta)$, but we do not make any assumption
on their sign.

The two phases $\varphi$ and $\widetilde{\varphi}$ depend on the Higgs 
background. Therefore, in the evaluation of the derivatives of $\veff$,
their contribution should not be neglected. To simplify the calculations,
we choose to redefine the fields in such a way that the top and stop mass 
matrices become real: 
\be
\label{redef}
\displaystyle{
t_L^{\, \prime} = e^{-i\frac{\varphi}{2}} t_L\,,\hspace{0.5cm}
t_R^{\, \prime} = e^{\,i\frac{\varphi}{2}} t_R\,,\hspace{1cm}
\tilde{t}_L^{\, \prime} = 
e^{-i\frac{\widetilde{\varphi}}{2}} \tilde{t}_L\,,\hspace{0.5cm}
\tilde{t}_R^{\, \prime} =
e^{\,i\frac{\widetilde{\varphi}}{2}} \tilde{t}_R \, . }
\ee
This redefinition allows us to combine $t_L$ and $t_R$, in the usual way,
into a four--component Dirac spinor, with a real field--dependent mass 
$m_t \equiv h_t \, |H_2^0|$. Moreover, the field--dependent stop mass 
matrix is now real and symmetric, thus it can be diagonalized by the 
orthogonal transformation:
\be
\left(
\begin{array}{c}
\widetilde{t}_1 \\ \widetilde{t}_2
\end{array}
\right)
\; = \;
\left(
\begin{array}{cc}
\cos\ths & \sin\ths\\
-\sin \ths & \cos\ths 
\end{array}
\right)\;
\left(
\begin{array}{c}
\widetilde{t}_L \\ \widetilde{t}_R
\end{array}
\right) \, .
\ee
The field--dependent stop masses are: 
\be
m^2_{\tilde{t}_{1,2}} = \frac{1}{2} \left[ ( m_L^2 + m_R^2) \pm 
 \sqrt{ (m_L^2-m_R^2)^2 + 4 \, |\mix|^2} \right] \, ,
\label{mstop}
\ee
and the mixing angle $\ths$ is also a field--dependent quantity, 
defined by:
\be
\sin 2\ths = \frac{ 2 \, |\mix|}{\msqu - \msqd} \, .
\label{sint}
\ee
Notice that, in this case, $0 \leq \ths < \pi/2$, in contrast with 
the usual field--independent definition for the angle $\thz$ that 
diagonalizes the stop mass matrix at the minimum,
\be
\label{sinmin}
\sin 2 \thz = \frac{2\,m_t\,(A_t + \mu\cot\beta)}{\diff}\,,
\ee
which leads to $-\pi/2 \leq \thz < \pi/2$.
The redefinition (\ref{redef}) has no effect on almost all 
field--dependent interaction vertices, with the only 
exception of the ones involving top, stop and gluino,
which acquire a dependence on the phase difference 
$(\varphi - \widetilde{\varphi})$:
\be
\displaystyle{
{\cal L}_{t \tilde{t} \tilde{g}} =
- \sq2 \, g_s  \, \left( 
\overline{t_L} \,\tilde{g} \: T \: \tilde{t}_L \,
e^{\frac{i}{2}(\varphi - \widetilde{\varphi})}
-
\overline{t_R} \,\tilde{g}\: T \: \tilde{t}_R \,
e^{\frac{i}{2}(\widetilde{\varphi} - \varphi)} 
 \right) + {\rm h.c.} \, , }
\label{gint}
\ee
where  $T$ are the $SU(3)$ generators in the fundamental 
representation and all color indices are understood.

Before presenting the results for the ${\cal O}(\at \as)$ corrections
to the neutral Higgs boson masses, we discuss the general structure of 
the pure SQCD corrections to the one--loop ${\cal O}(\at)$ results, 
namely the ${\cal O}(\at \as^n)$ terms for generic $n >0$. For the
computation of this class of corrections, the effective potential 
can be expressed as a function of five field--dependent quantities, 
which can be chosen as follows.
The masses $m_t^2 = h_t^2 |H_2^0|^2 \,$, $\msqu$ and $\msqd$, 
the last two as defined in eq.~(\ref{mstop}). The mixing parameter
$\c2t2 \equiv 1 - \sin^2 2\ths$, where $\sin 2\ths$ is given in 
eq.~(\ref{sint}). Finally, a parameter that, according to 
eq.~(\ref{gint}), should be a function of the phase difference 
$\varphi - \widetilde{\varphi}$: we conveniently choose it as 
$\cpp \equiv \cos\,(\varphi - \tilde{\varphi})$, where
\be
\cos\,(\varphi - \tilde{\varphi}) = 
\frac{ {\rm Re}(\mixt)\,{\rm Re}(\mix) + {\rm Im}(\mixt)\,{\rm Im}(\mix)}
{|\mixt| \, |\mix|} \, ,
\ee
$\mixt$ and $\mix$ are defined in eqs. (\ref{mixt}) and (\ref{mix}), 
respectively, and $\langle \cpp \rangle = \pm 1$. A sixth parameter, 
the gluino mass $\mgl$, does not depend on the Higgs background:
we can restrict ourselves to positive values of $\mgl$ if we 
neglect possible CP--violating phases and we allow for arbitrary
signs of $\mu$ and $A_t$.

According to eq. (\ref{msij2}), in the limit of neglecting the $SU(2)_L
\times U(1)_Y$ gauge contributions beyond the tree level, the radiative 
corrections to the neutral CP--even Higgs boson mass matrix can be 
parametrized as:
\be
\left( {\cal M}^2_S \right)_{ij}  
= 
\left( {\cal M}^2_S \right)^0_{ij} 
+
\left(\Delta{\cal M}^2_S\right)_{ij}
\,,
\ee
where
\be
\label{mszero}
\left( {\cal M}^2_S \right)^0_{ij} 
=
(-1)^{i+j} \left[\;\left( {\cal M}^2_P \right)_{ij}  
+
{(g^2 + g'^2) \over 2} v_i v_j\right] 
\ee
is fully determined by the input parameters $m_Z$, $m_A$ and 
$\tan \beta$ since, at ${\cal O}(\as \at)$, $m_Z^2=(g^2 + g'^2)
(v_1^2+v_2^2)/4$. The corrections that have not been reabsorbed 
in $({\cal M}^2_S)^0$ are contained in: 
\be
\label{Dms}
\left(\Delta{\cal M}^2_S\right)_{ij}  = 
-(-1)^{i+j}\left.
\frac{\partial^2 V}{\partial P_i \partial P_j}\right|_{\rm min}
+\left.\frac{\partial^2 V}{\partial S_i \partial S_j}\right|_{\rm min}~.
\ee
Exploiting the field--dependence of the parameters, a wise 
although lengthy application of the chain rule for the 
derivatives of the effective potential leads to:

\bea
\label{dms11}
\hspace{-1cm}\left(\Delta {\cal M}^2_S\right)_{11} & = &
\frac{1}{2} \, h_t^2 \,\mu^2 \,\s2t^2  \,F_3 \, , \\
\label{dms12}
\hspace{-1cm}\left(\Delta {\cal M}^2_S\right)_{12} & = &
h_t^2 \,\mu\, m_t\, \s2t \,  F_2 
+ \frac{1}{2}\, h_t^2\, A_t \,\mu \, \s2t^2 \, \left(F_3 
+ \Delta F_3 \right) \, , \\
\label{dms22}
\hspace{-1cm}\left(\Delta {\cal M}^2_S\right)_{22} & = &
2\, h_t^2\, m_t^2\, F_1
+ 2\, h_t^2\, A_t\, m_t\, \s2t\, 
\left( F_2 + \Delta F_2 \right)
+ \frac{1}{2}\, h_t^2\, A_t^2\, \s2t^2\, 
\left(F_3 + 2\, \Delta F_3 \right) \, ,
\eea
where $\s2t \equiv \sin 2\thz$ refers to the stop mixing angle defined 
in the usual field--independent way [see eq.~(\ref{sinmin})]. 
The functions $F_i$ ($i=1,2,3$) can be decomposed as
$F_i = \widetilde{F}_i + \Delta  \widetilde{F}_i$, where the
$\Delta \widetilde{F}_i$ include the renormalization of the
common factors multiplying $F_i$  (i.e. $h_t^2,\,\mt,\,\s2t$), 
and
\bea
\widetilde{F}_1 & = &
\frac{\partial^{\,2} V}{(\partial m_t^2)^2} 
+ \frac{\partial^{\,2} V}{(\partial \msqu)^2} 
+ \frac{\partial^{\,2} V}{(\partial \msqd)^2} \nn\\
&&
+ 2\, \frac{\partial^{\,2} V}{\partial m_t^2 \partial \msqu}
+ 2\, \frac{\partial^{\,2} V}{\partial m_t^2 \partial \msqd}
+ 2\, \frac{\partial^{\,2} V}
{\partial \msqu\partial \msqd}
+ \frac{1}{4 \, m_t^4} \frac{\partial V}{\partial \cpp} \, ,
\\&&\nn\\
\widetilde{F}_2 & = &
\frac{\partial^{\,2} V}{(\partial \msqu)^2}
- \frac{\partial^{\,2} V}{(\partial \msqd)^2}
+ \frac{\partial^{\,2} V}{\partial m_t^2 \partial \msqu}
- \frac{\partial^{\,2} V}{\partial m_t^2 \partial \msqd} 
- \frac{(\s2t)^{-2}}{m_t^2 \, (\msqu-\msqd)}\,
\frac{\partial V}{\partial \cpp}\nn\\
&&
- \frac{4 \, \csenza}{\msqu-\msqd}
\left( 
\frac{\partial^{\,2} V}{\partial \c2t2 \partial m_t^2}
+ \frac{\partial^{\,2} V}{\partial \c2t2 \,\partial \msqu}
+ \frac{\partial^{\,2} V}{\partial \c2t2 \,\partial \msqd}
\right) \, ,
\\&&\nn\\
\widetilde{F}_3 & = &
\frac{\partial^{\,2} V}{(\partial \msqu)^2}
+ \frac{\partial^{\,2} V}{(\partial \msqd)^2}
- 2\, \frac{\partial^{\,2} V}
{\partial \msqu\partial \msqd}
- \frac{2}{\msqu-\msqd}
\left( \frac{\partial V}{\partial \msqu}
- \frac{\partial V}{\partial \msqd} \right)
+ \frac{ 4\, (\s2t)^{-4}}{(\msqu-\msqd)^2}\,
\frac{\partial V}{\partial \cpp}\nn\\
&&
+ \frac{ 16\, \csenza}{(\msqu-\msqd)^2}\,
\left( \csenza \frac{\partial^{\,2} V}{(\partial \c2t2)^2}
+ 2 \, \frac{\partial V}{\partial \c2t2} \right)
- \frac{ 8\, \csenza}{\msqu-\msqd}\,
\left( \frac{\partial^{\,2} V}{\partial \c2t2 \,\partial \msqu}
- \frac{\partial^{\,2} V}{\partial \c2t2 \,\partial \msqd} \right)~.
\eea
In the above equations, $\csenza = 1 - \s2t^2$ and the derivatives are 
evaluated at the minimum of $\veff$. 
As can be seen from  eqs.~(\ref{dms11})--(\ref{dms22}), 
in every entry of $\Delta {\cal M}^2_S$ the
$F_i$ terms are multiplied by different combinations of $\mu$ and
$A_t$. These two parameters  do not renormalize in the same way,
thus the contributions induced by their renormalization 
cannot be absorbed into the $F_i$, but should be separately 
taken into account. However, since at ${\cal O}(\as)\;$
$\mu$  does not renormalize,
we have inserted in eqs.~(\ref{dms12})--(\ref{dms22}) only the two
factors $ \Delta F_2$ and  $\Delta F_3$ that take into account the 
renormalization of $A_t$.

To evaluate the functions $(F_1,F_2,F_3)$, two strategies come to mind:
i) evaluate explicitly the effective potential and then  differentiate 
with respect to the relevant field--dependent quantities; ii) use a
well--known fact, that the derivative of a bubble diagram with respect 
to an internal mass is still a diagram of the same kind, to compute 
directly the derivatives of the effective potential, without evaluating 
the effective potential itself. In our calculation of the ${\cal O}
(\at \as)$ corrections to the neutral Higgs boson masses we followed 
the latter strategy. The corresponding corrections to the effective 
potential can be found in eq.~(4) of the second paper of ref.~\cite{ez}, 
with the understanding that the last two terms, proportional to $\mgl 
\mt \s2t$, should be multiplied by $\cpp$. However, only the second 
derivatives of $\cpp$ with respect to the fields $P_i$ are different 
from zero at the minimum of the potential. Therefore, this extra term 
does not contribute to the expression for $m_h$ in the decoupling limit 
of very large $m_A$, where the results of that paper are applicable.

We give now the explicit expressions for the ${\cal O}(\at \as)$ 
contribution to the functions $(F_1,F_2,F_3)$. For completeness, 
we recall first the one--loop result \cite{erz2}:
\be
\label{fi1l}
F_1^{\rm 1\ell} = \frac{N_c}{16\,\pi^2}\ln\frac{\msqu \msqd}{m_t^4}\,,
\hspace{0.7cm}
F_2^{\rm 1\ell} = \frac{N_c}{16\,\pi^2}\ln\frac{\msqu}{ \msqd}\,,
\hspace{0.7cm}
F_3^{\rm 1\ell} = \frac{N_c}{16\,\pi^2} \left( 2 -
\frac{\msqu+\msqd}{\msqu-\msqd}
\,\ln\frac{\msqu}{ \msqd} \right) \, ,
\ee
where $N_c = 3$ is a color factor. We assume that the  ${\cal O}(\at)$
one--loop contribution is written in terms of top and stop parameters
evaluated in the $\overline{\rm DR}$ scheme [$v_1$ and $v_2$ are 
automatically defined in the $\ov{\rm DR}$ scheme by eqs.~(\ref{minima1}) 
and (\ref{minima2}), and the same is true for $\tan \beta=v_2/v_1$]. 
The two--loop ${\cal O}(\at \as)$ contributions to the functions 
$(F_1,F_2,F_3)$, in units of $g_s^2\,C_F\,N_c/(16\,\pi^2)^2$ 
(where $C_F = 4/3$), and in the $\overline{\rm DR}$ renormalization scheme
(here and hereafter $\overline{\rm DR}$ quantities will be denoted
by a hat), are given by: 
\bea
\label{f12l}
 \hat{F}_{1}^{\rm 2\ell} & = &
- 6\,\left(1-\ln \frac{m_t^2}{Q^2}\right)
+ 5\,\ln \frac{\msqu \msqd}{m_t^4}
+ \ln^2\frac{\msqu \msqd}{m_t^4}
+ 8\,\myltq\nn \\ 
\hspace{-2cm}&&\nn\\
\hspace{-2cm}&&
-4\,\left(\myltuq +\myltdq\right)
-\, c^2_{2 \theta}\,\left[ 
2  - \ln\frac{\msqu\,\msqd}{Q^4} - \ln^2\frac{\msqu}{\msqd} 
\right]\nn\\
\hspace{-2cm}&&\nn\\
\hspace{-2cm}&&
-\, s^2_{2 \theta}\,\left[  
\frac{\msqu}{\msqd} \,\left(1 - \myltu \right)
+ \frac{\msqd}{\msqu} \,\left(1 - \myltd \right)
\right]\nn\\
\hspace{-2cm}&&\nn\\
\hspace{-2cm}&&
+ \; f_1(m_t\,,\mgl\,,\msqu\,,\msqd\,,\s2t\,,Q)\; 
+ \; f_1(m_t\,,\mgl\,,\msqd\,,\msqu\,,-\s2t\,,Q)~,\\
\hspace{-2cm}&&\nn\\
\hspace{-2cm}&&\nn\\
\label{f22l}
\hat{F}_{2}^{\rm 2\ell} & = &
5\,\ln\frac{\msqu}{\msqd}
- 3\,\left(\myltuq - \myltdq\right) 
+ \,c^2_{2 \theta}\,
\left[ 5\,\ln\frac{\msqu}{\msqd}  
\right.\nn\\ 
\hspace{-2cm}&&\nn\\
\hspace{-2cm}&& \left.
- \frac{\msqu + \msqd}{\msqu - \msqd}
\,\ln^2\frac{\msqu}{\msqd}
- \frac{2}{\diff}\,\left(\msqu\,\myltu - \msqd\,\myltd\right)
\,\ln\frac{\msqu}{\msqd} \,\right] \nn\\ 
\hspace{-2cm}&&\nn\\
\hspace{-2cm}&&
+ \,s^2_{2 \theta}\,
\left[ 
\frac{\msqu}{\msqd} \,\left(1 - \myltu \right)
- \frac{\msqd}{\msqu} \,\left(1 - \myltd \right)
\right] \nn\\
\hspace{-2cm}&&\nn\\
\hspace{-2cm}&&
+ \; f_2(m_t\,,\mgl\,,\msqu\,,\msqd\,,\s2t\,,Q)\; 
- \; f_2(m_t\,,\mgl\,,\msqd\,,\msqu\,,-\s2t\,,Q)~,\\
\hspace{-2cm}&&\nn\\
\hspace{-2cm}&&\nn\\
\label{f32l}
 \hat{F}_{3}^{\rm 2\ell} & = & 
\frac{16 \pi^2}{N_c} F_3^{\rm 1\ell} \,(3 + 9 \,c^2_{2 \theta})
+ 4 
-\frac{3 + 13\,\csenza}{\diff}\,
\left(\msqu\,\myltu - \msqd\,\myltd\right)\nn \\ 
\hspace{-2cm}&&\nn\\
\hspace{-2cm}&&
+ \,3\,\frac{\msqu +\msqd}{\msqu - \msqd}\,
\left( \myltuq - \myltdq \right)  
- \csenza\, \left[ 
4 - \left(\frac{\msqu+\msqd}{\diff}\right)^2\,\ln^2\frac{\msqu}{\msqd}
\right.\nn\\
\hspace{-2cm}&&\nn\\
\hspace{-2cm}&& \left. 
- 6\,\frac{\msqu+\msqd}{(\diff)^2}\,
\left(\msqu\,\myltu - \msqd\,\myltd\right)\,\ln\frac{\msqu}{\msqd}\right]
- \,s^2_{2 \theta}\,\left[
\frac{\msqu}{\msqd} + \frac{\msqd}{\msqu}\right.\nn\\
\hspace{-2cm}&&\nn\\
\hspace{-2cm}&& 
\left.
+ 2\,\ln\frac{\msqu\,\msqd}{Q^4}
- \frac{\msquq}{\msqd\,(\diff)}\,\myltu
+ \frac{\msqdq}{\msqu\,(\diff)}\,\myltd\,
\right]\nn\\
\hspace{-2cm}&&\nn\\
\hspace{-2cm}&&
+ \; f_3(m_t\,,\mgl\,,\msqu\,,\msqd\,,\s2t\,,Q)\; 
+ \; f_3(m_t\,,\mgl\,,\msqd\,,\msqu\,,-\s2t\,,Q)~.
\eea
In the above expressions, $Q$ indicates the $\ov{{\rm DR}}$
renormalization scale, and
the functions $f_i\,$ contain contributions coming from the 
top-stop-gluino diagrams (fig. \ref{figdiag}d): their explicit 
expressions are presented in appendix A.

To obtain the  ${\cal O}(\at \as)$ corrections to the Higgs mass 
entries, we also need explicit expressions for the $\Delta F_i$ terms 
of eqs.~(\ref{dms12})--(\ref{dms22}). In the $\ov{\rm DR}$ scheme, 
and in units of $g_s^2\,C_F\,N_c/(16\,\pi^2)^2$, they are:
\bea
\Delta \hat{F}_2 & = & \frac{2\, \mgl}{A_t}
\left(\ln^2\frac{\msqd}{Q^2} - \ln^2\frac{\msqu}{Q^2}\right)
\, , \\
&&\nn\\
\Delta \hat{F}_3 & = & \frac{\mgl}{A_t} \left[
8 - 2\,\frac{\msqu+\msqd}{\msqu-\msqd}\,
\left(\ln^2\frac{\msqd}{Q^2} - \ln^2\frac{\msqu}{Q^2}\right) \right. \nn\\
&& + \left. \frac{8}{\msqu-\msqd}\,
\left(\msqd\,\ln\frac{\msqd}{Q^2} - \msqu\,\ln\frac{\msqu}{Q^2}\right)
\right] \, .
\label{df3}
\eea

A comment on eqs. (\ref{f12l})--(\ref{df3}) is in order. 
These equations show an explicit dependence on the renormalization 
scale $Q$, connected with our choice of expressing the top and stop 
parameters in the $\overline{\rm DR}$ scheme. This dependence is
cancelled by the implicit dependence of the $\ov{\rm DR}$
parameters, so that the entries of  $\Delta {\cal M}^2_S$
are scale--independent. This fact becomes manifest if we reexpress
the top and stop parameters in a physical scheme such as the on--shell 
(OS) scheme. To ensure this scale--independence,
it was crucial to include the contributions induced by $\partial V /
\partial \cpp$.  If these terms were  neglected, and the limit
$m_A \rightarrow \infty$ were taken, one would still find a 
scale--independent result for the ${\cal O}(\at\,\as)$ corrections to 
$m_h$ (thanks to the fact that, in such a limit, $m_h$ does not 
depend upon $m_A$), but not for the corrections to $m_H$. 
It may be useful to recall that, because $v_1$ and $v_2$ are automatically
defined in the $\ov{\rm DR}$ scheme, $\left({\cal M}_S^2\right)^0$ has 
an implicit scale dependence, since 
${\cal M}_P^2$ in eq.~(\ref{mszero}) contains $\tan\beta$. This 
residual scale dependence could be removed by including the 
momentum--dependent parts of the self--energies in the two--loop 
computation.

To obtain the ${\cal O}(\at \as)$ corrections in some other renormalization
scheme, $R$, we just need to shift the top and stop parameters appearing in 
the one--loop term. Indicating the general mass in the $\overline{\rm DR}$ 
scheme as $m^{\overline{\rm DR}}$, and the same quantity in the $R$
scheme as $m$, we can write the one--loop relation $m = m^{\ov{\rm DR}} 
- \delta m$. Then, once the one--loop contribution is written in terms 
of $R$ quantities, the two--loop ${\cal O}(\at \as)$ corrections in 
the $R$ scheme  can be obtained through:
\bea 
F_1^{\rm 2\ell} & = & \hat{F}_1^{\rm 2\ell} 
+\frac{N_c}{16\,\pi^2}\,
\left(\frac{\delta\msqu}{\msqu} + \frac{\delta\msqd}{\msqd} 
- 4\, \frac{\delta m_t}{m_t} \right)
+ 4\, \frac{\delta m_t}{m_t}\, F_1^{\rm 1\ell} \,
\label{f12lr}\\
F_2^{\rm 2\ell} & = & \hat{F}_2^{\rm 2\ell} 
+\frac{N_c}{16\,\pi^2}\,
   \left(\frac{\delta\msqu}{\msqu} - \frac{\delta\msqd}{\msqd}\right) 
+ \left( 3\, \frac{\delta m_t}{m_t} + \frac{\delta \s2t}{\s2t} \right)
\, F_2^{\rm 1\ell} \, ,
\label{f22lr}\\
F_3^{\rm 2\ell} & = & \hat{F}_3^{\rm 2\ell}
+ \frac{N_c}{16\,\pi^2}\,
\left(2 \,\frac{\msqu \msqd}{(\msqu-\msqd)^2}\ln\frac{\msqu}{\msqd}
- \frac{\msqu+\msqd}{\msqu-\msqd}\right)\;
\left(\frac{\delta\msqu}{\msqu} - \frac{\delta\msqd}{\msqd} \right)\nn \\
&& \hspace{0.7cm} + \left(2\,\frac{\delta m_t}{m_t} 
+ 2\,\frac{\delta \s2t}{\s2t} \right)\, F_3^{\rm 1\ell}
\label{f32lr} \, , \\
&&\nn\\
(\Delta F_2) & = & (\Delta \hat{F}_2) + 
\frac{\delta A_t}{A_t} \, F_2^{\rm 1\ell} \, ,
\label{df222}\\
(\Delta F_3) & = & (\Delta \hat{F}_3) +
\frac{\delta A_t}{A_t} \, F_3^{\rm 1\ell} \, ,
\label{df312}
\label{df322}
\eea
where all the quantities that appear in 
eqs.~(\ref{f12lr}--\ref{df322}) are meant 
in the $R$ scheme. The explicit expressions that allow us 
to perform the one--loop shift for ($m_t,\, \msqu,\, \msqd,\, 
\s2t,\,A_t$), from the $\ov{\rm DR}$ to the OS renormalization
scheme, are listed in appendix B.


\section{Discussion}
Using the formalism of the effective potential, we have obtained 
complete, explicit, analytical expressions for the 
${\cal O}(\at \as)$ two--loop
corrections to the MSSM mass matrix for the CP--even Higgs bosons.
Our input parameters are: ($m_Z,\,m_A,\,\tan\beta$), already
appearing in the tree-level result; the parameters of the top and 
stop sectors, appearing in the ${\cal O}(\at)$ one--loop correction,
for example $(m_t,\,\msqu,\,\msqd,\,\s2t,\,\mu)$;
the gluino mass and the strong coupling constant, appearing only
at the two--loop level. We have presented our results in such a
way that the input parameters of the top and stop sectors can
be given either in the $\ov{{\rm DR}}$ scheme or in some version
of the on--shell scheme. Also, we have included in appendix~C 
the ${\cal O}(\at \as)$ corrections to the relation that gives 
$m_A^2$ in terms of $m_3^2$ and $\tan \beta$. This result can be 
useful if one deals with models that predict the low--energy values 
of the soft supersymmetry--breaking parameters. 

Our effective potential calculation is equivalent to the
evaluation of the Higgs self--energies in the limit of 
vanishing external momentum. A diagrammatic computation of the two--loop
${\cal O}(\at\as)$ contributions to the Higgs boson self--energies at zero
external momentum has been performed in \cite{hhw2l}. Analytical 
formulae, valid in the simplified case of degenerate soft stop masses 
and zero mixing (with $\mu = A_t = 0$), have been presented in the first 
paper of ref.~\cite{hhw2l}. For arbitrary values of the top and stop 
parameters, however, the complete analytical result of \cite{hhw2l}
is far too long to be explicitly presented, and is only available 
as a computer code \cite{FeyH}. We have checked that, in the case of zero 
mixing and degenerate stop masses, our results coincide with those of 
\cite{hhw2l}. Moreover, after taking into account the difference in the 
definitions of the on--shell renormalized angle $\thz$ (see appendix B), 
we find perfect agreement with the numerical results of \cite{FeyH}, for 
arbitrary values of all the input parameters. 

A calculation of both ${\cal O}(\at\,\as)$ and ${\cal O}(\at^2)$ 
two--loop corrections to the lightest Higgs boson mass $m_h$, based on 
the formalism of the effective potential, has been presented in \cite{ez}.
In these papers, however, the dependence of the stop masses 
and mixing angles on the fields $P_i$ (the CP--odd components of the 
neutral Higgs fields) is not taken into consideration. Therefore, 
the ${\cal O}(\at\,\as)$ corrections to the input parameter $m_A$ 
are not evaluated.
If one wants to relate the input parameters to measurable
quantities, the computation is applicable only in the limit 
$m_A \gg m_Z$, in which $m_h$ is nearly independent of $m_A$. 
However, the results of [15] allow to express $m_h$ as a function of
the (unphysical) renormalized parameter $m_3^2$ in the $\ov{{\rm DR}}$ scheme.
Moreover, while an analytical formula for $V_{\rm eff}$ is given (in terms
of $m_3^2$ and of the fields $S_i$), the results of \cite{ez} 
for $m_h$ are available in 
numerical form, and simple analytical formulae are provided only in 
the case of universal soft stop masses and small stop mixing. 

The corrections controlled by $\as$ introduce a new mass scale in the 
prediction of the MSSM neutral Higgs boson masses, namely the gluino mass 
$\mgl$. To avoid dealing with many different scales, $\mgl$ is usually set 
to be of the same magnitude of the stop masses. Notice that, in a scenario
where stops and gluinos are all heavy and approximately degenerate, 
with masses ${\cal O}(M_S)$, only the function $\hat{F}_{1}$ contains  
large logarithms of the ratio $\mt/ M_S$. Instead, as we have explicitly
checked, $\hat{F}_{2}$ and  $\hat{F}_{3}$ are finite in the limit 
$\mt \rightarrow 0$. They contribute only to the matching conditions 
between the MSSM
and the effective theory below the scale $M_S$, to be identified at
${\cal O} (\at \as)$ with a two--Higgs--doublet version of the Standard
Model. However, if the term $A_t + \mu \cot \beta$ is very large, or 
$m_L$ and $m_R$ are very different, or both, the two stop mass eigenstates 
can have a wide mass gap, and large logarithms of the ratio $\mtu/\mtd$
can then be present in all the $\hat{F}_{i}$ terms. It should be 
mentioned that in this case the low-energy effective theory is  
different from a two--Higgs--doublet version of the SM, and indeed
much more complicated already in the case of small stop mixing 
\cite{espnav}, not to mention the difficult case of large stop mixing.

We can also envisage a scenario in which the gluino is much heavier
than the top and the stops. Eqs.~(\ref{f1picc})--(\ref{f3picc}) 
contain terms proportional to powers of $\mgl$ that can be 
potentially large. This powerlike behavior is actually cancelled 
in the OS schemes by the finite parts of the relevant shifts. However,
as already noticed in \cite{hhw2l}, the gluino does not fully 
decouple, and $m_h$ increases logarithmically with $\mgl$ when 
the latter becomes very large. On the other hand, it must be 
noticed that, in the $\overline{\rm DR}$ scheme,
 some terms proportional to $\mgl$ and $\mgl^2$ are not cancelled, 
and in the limit of heavy gluino the two--loop corrections to the 
Higgs masses can become very large : this is
related with the non--decoupling properties of mass--independent
renormalization schemes such as $\ov{{\rm DR}}$. 

From eqs.~(\ref{f1picc})--(\ref{f3picc}), we can derive approximate 
formulae for the two--loop corrections to ${\cal M}^2_S$ 
in the case of large gluino mass, keeping the leading terms in an 
expansion of the complete result in powers of $\mgl$.
Doing so, the asymptotic behavior is approached quite slowly as
$\mgl$ increases. The reason is that some terms that are formally 
suppressed by inverse powers of $\mgl$ are indeed enhanced by large 
numerical factors. In order to get an accurate approximation to the 
correct result, it is then preferable to include also
the next--to--leading terms in the expansion in powers of $\mgl$. 
Specializing for simplicity to the case of degenerate soft stop masses
and $\mu = A_t = 0$, so that $\msqu = \msqd \equiv \msqc$ and the only 
non--zero correction to the Higgs mass matrix is 
$\left(\Delta {\cal M}^2_S\right)_{22}$, we find:
\bea 
\hspace{-1.5cm}\left(\Delta {\cal M}^2_S\right)_{22} & = &
\frac{h_t^2 \, g_s^2}{8 \,\pi^4}\,m_t^2\,\left(
\frac{2\,\pi^2}{3} - 1 
- 6\, \ln\frac{\mgl^2}{m_t^2}
- 3\, \ln^2\frac{\msqc}{m_t^2}
+ 2\, \ln^2\frac{\mgl^2}{\msqc} \right)\hspace{2cm}\nn
\eea
\vskip -0.5cm
\bea
\hspace{-1.5cm}&&
+\frac{h_t^2 \, g_s^2}{64 \,\pi^4}\,\frac{m_t^2}{m_g^2}\,\left(
\frac{32\,\pi^2}{3}\,(2\,m_t^2 + \msqc) 
-\frac{160\,m_t^2 + 112\,\msqc}{3}
-\frac{352}{3}\,m_t^2\,\ln\frac{\mgl^2}{m_t^2}
+32\,m_t^2\,\ln^2\frac{\mgl^2}{m_t^2}
\right.\nn\\ 
\hspace{-1.5cm}&&\nn\\
\hspace{-1.5cm}&&\left.
\label{hgl2}
-\frac{224\,m_t^2 + 288 \,\msqc}{3}\,\ln\frac{\mgl^2}{\msqc}
+32\,(m_t^2 + \msqc)\,\ln^2\frac{\mgl^2}{\msqc}
-32\,m_t^2\,\ln^2\frac{\msqc}{m_t^2}\,\right)
\;\;\; + \;{\cal O}\left(\mgl^{-4}\right)\;.
\eea
In deriving eq.~(\ref{hgl2}), we have assumed an on--shell renormalization 
for the top and stop parameters. As anticipated above, if we were to write 
the one--loop corrections in terms of $\ov{{\rm DR}}$ parameters, we would 
find in $\Delta {\cal M}_S^2$ terms proportional to $\mgl^2$. In fact, as 
can be seen from eq.~(\ref{dmt1}), the finite shift $\delta\msqc$ scales 
for large $\mgl$ as $\mgl^2 \,(\ln (\mgl^2/Q^2) -1)$, and cancels in 
the OS scheme similar terms present in the $f_i$.

\begin{figure}[t!]
\begin{center}
\mbox{
\hspace{-.4cm}
\epsfig{figure=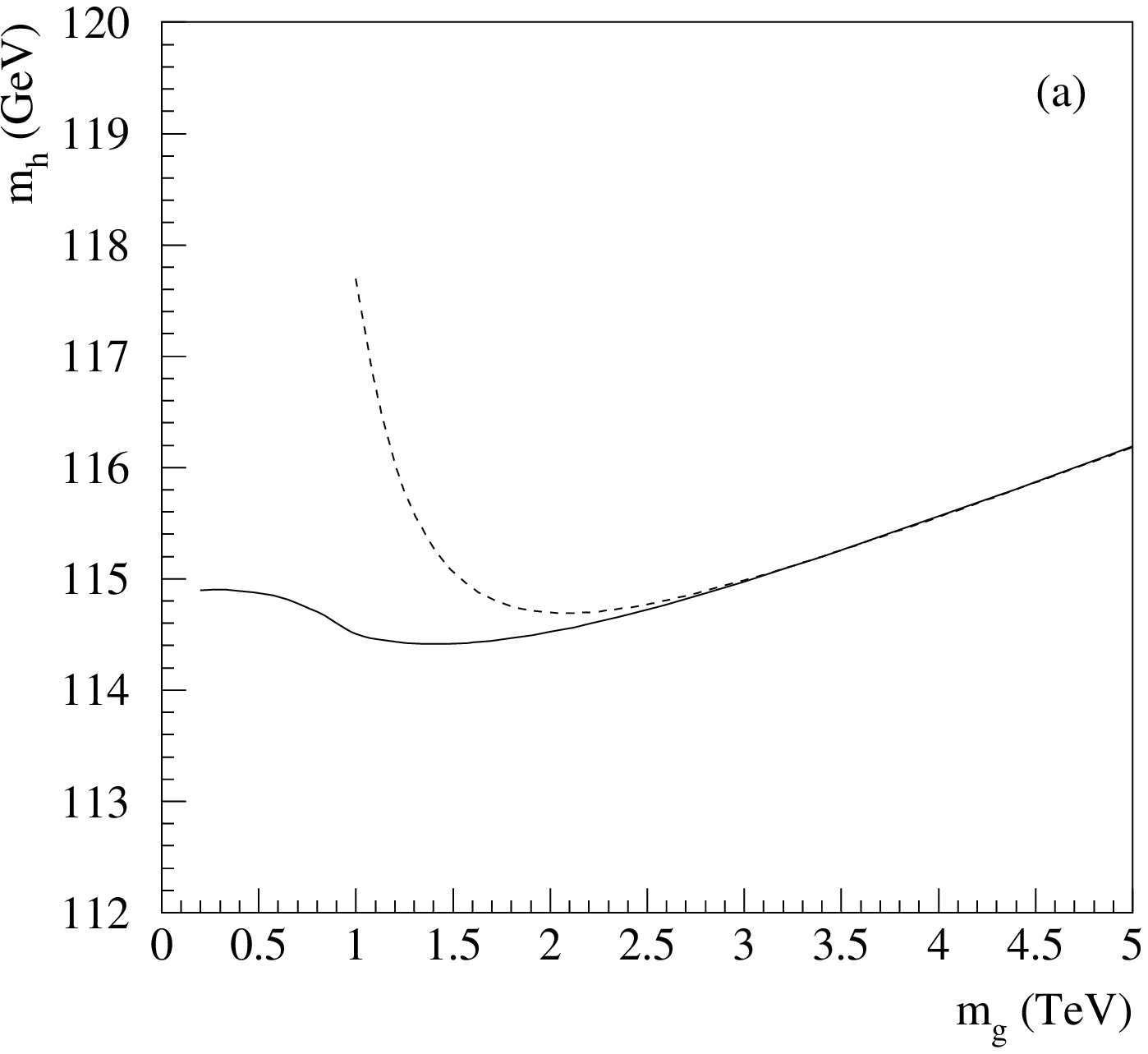,width=9cm,height=8cm}
\hspace{-1cm}
\epsfig{figure=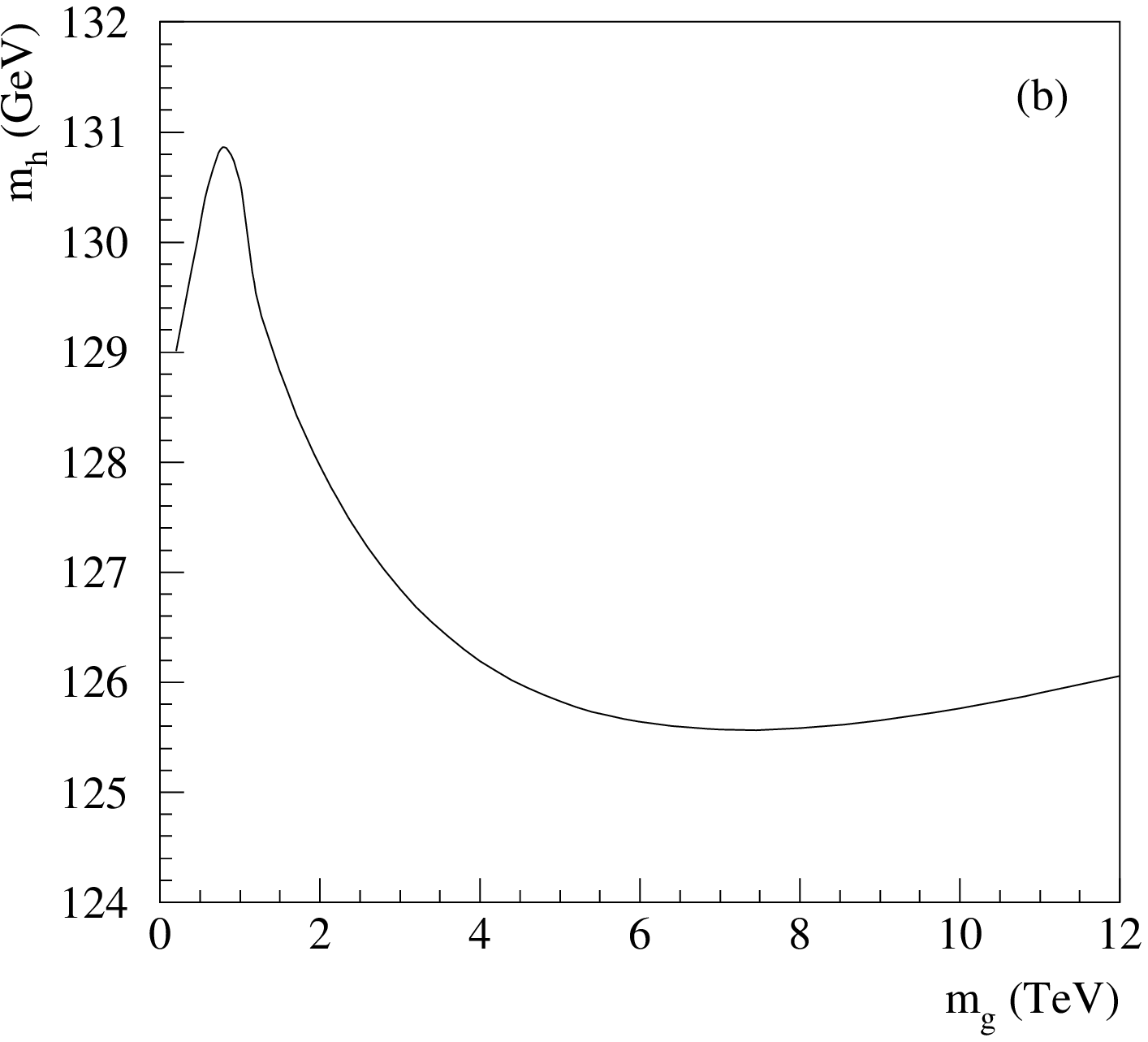,width=9cm,height=8cm}}
\end{center}
\vspace{-0.5cm}
\caption{The mass $m_h$ from our effective potential calculation, as a 
function of the gluino mass $\mgl$, for (a) no mixing 
($\mtu = \mtd = 1015$ GeV, $\,\s2t=0$) 
and (b) large mixing 
($\mtu = 1175$ GeV, $\,\mtd = 825$ GeV, $\,\s2t=1$) 
in the stop sector. 
The other MSSM parameters are: $m_A = 500 $ GeV, $\tan\beta = 10,
\, \mu=0$ and $m_t=175$ GeV.  
The dashed line in fig. \ref{gluinofig}a is the approximate result of 
eq.~(\ref{hgl2}).}
\label{gluinofig}
\end{figure}

Some representative results for the heavy--gluino limit are shown 
in fig. \ref{gluinofig}. We have plotted $m_h$, as obtained 
from our complete formulae, as a function of the gluino mass. 
We employ on--shell top and stop parameters (for their definitions, 
see appendix B). We consider two examples, with degenerate 
soft stop masses and either no mixing or large mixing in the 
stop sector. The numerical inputs we use are: $m_A = 500 $ GeV, 
$\tan\beta = 10,\, \mu = 0,\, m_t = 175$ GeV, and the values
$\mtu = \mtd = 1015$ GeV, $\,\s2t = 0$ (no mixing, fig.~\ref{gluinofig}a) 
and $\mtu = 1175$ GeV, $\mtd = 825$ GeV, $\s2t = 1$ 
(large mixing, fig.~\ref{gluinofig}b). 
For the one--loop ${\cal O}(\at)$ corrections to 
${\cal M}^2_S$ we have used the  effective potential result of \cite{erz2}.
As can be seen from figure \ref{gluinofig}, $m_h$ reaches a maximum at low 
values of $\mgl$, decreases for intermediate values and then increases 
logarithmically when $\mgl$ becomes very large. Comparing the cases 
of no mixing  and large stop mixing, we see that in the latter case the peak 
of $m_h$ at small $\mgl$ is much more pronounced, and the asymptotic 
increase of  $m_h$ starts at higher values of $\mgl$. 
In fig.~\ref{gluinofig}a we have also plotted $m_h$ as obtained with the 
approximate formula of eq.~(\ref{hgl2}), valid in the case of no mixing. 
One can see that eq.~(\ref{hgl2}) approximates very well the complete 
result when $\mgl > 2\, m_{\tilde{t}}$.


\section*{Acknowledgements}
We would like to thank J.R.~Espinosa, F.~Feruglio, G.~Ganis
and especially A.~Brignole for discussions; the Physics Departments
of the Universities of Rome `La Sapienza' (G.D.) and Padua
(F.Z.) for their hospitality during part of this project.
This work was supported in part by the European Union under 
the contracts HPRN-CT-2000-00149 (Collider Physics) and
HPRN-CT-2000-00148 (Across the Energy Frontier).



\begin{appendletterA}
\section*{Appendix A: Analytical expressions for the gluino contributions }

The explicit expressions of the contributions to the functions
$\hat{F}_i^{\rm 2\ell}$ $(i=1,2,3)$ coming from the top--stop--gluino 
diagrams (fig. \ref{figdiag}d) are rather long but, as apparent from 
eqs.~(\ref{f12l})--(\ref{f32l}), they possess useful symmetry properties 
under the exchanges $\;\msqu \!\leftrightarrow \msqd\,,\;  \s2t 
\leftrightarrow -\s2t$. In units of $g_s^2\,C_F\,N_c/(16\,\pi^2)^2\,$,
where $C_F=4/3$ and $N_c=3$ are color factors, the explicit expressions 
of the functions $f_i$ ($i=1,2,3$) are:


\bea
\hspace{-1.5cm} f_1(m_t\,,\mgl\,,\msqu\,,\msqd\,,\s2t\,,Q) & = &  
4\,\frac{m_t^2+\mgl^2-\,\mgl\,m_t\,\s2t}{\msqu} \,
\left(1 - \mylg\right)
+ 4\, \ln\frac{m_t^2}{\mgl^2}\nn 
\eea
\vskip -0.5cm
\bea
\hspace{-1.5cm} && 
- 2\, \ln\frac{\msqu}{\mgl^2}
\;+\; \frac{2}{\den}\,\left[
4\,\mgl^4\,\ln\frac{\msqu}{\mgl^2}
+ \left(\mgl^4 - \msquq +m_t^2\,\left( 10\,\mgl^2 + 3\,m_t^2 
 + 2\,\frac{m_t^2\,\mgl^2-m_t^4}{\msqu}\right)\right)
\,\ln\frac{m_t^2}{\mgl^2}\,\right]\nn\\
\hspace{-1.5cm}&&\nn\\
\hspace{-1.5cm} && 
+ \frac{2\,\mgl\,\s2t}{m_t}\,
\left(\myltuq + 2\,\mylt\,\myltu\right)
+ \frac{4\,\mgl\,\s2t}{m_t\,\den}\,\left[
\mgl^2\,\left(\msqu-m_t^2-\mgl^2\right)
\,\ln\frac{\msqu}{\mgl^2} \right.\nn\\
\hspace{-1.5cm}&&\nn\\
\hspace{-1.5cm} && \hspace{3cm} \left.
+ \,m_t^2 \left(\msqu - 3\,\mgl^2 - 2 \,m_t^2 
-\frac{m_t^2\,\mgl^2- m_t^4}{\msqu}\right)
\,\ln\frac{m_t^2}{\mgl^2}\,\right]\nn\\
\hspace{-1.5cm}&&\nn\\
\hspace{-1.5cm} && + \left[ 
\frac{4\,\mgl^2\,\left(m_t^2+\mgl^2-\msqu - 2\,\mgl\,m_t\,\s2t\right)}
{\den}
- \frac{4\,\mgl\,\s2t}{m_t}\right]\,
\Phi\,(m_t^2,\msqu,\mgl^2) \, , \hspace{3cm}
\label{f1picc}
\eea

\vskip 0.5cm


\bea
\hspace{-1.5cm}f_2(m_t\,,\mgl\,,\msqu\,,\msqd\,,\s2t\,,Q) & = &
4\, \frac{m_t^2 + \mgl^2}{\msqu}
-\frac{4\,\mgl\,\s2t}{m_t\,(\diff)}\,\left(
3 \,\msqu - \frac{m_t^2\,\msqd}{\msqu} \right)\hspace{1cm}\nn 
\eea
\vskip -0.5cm
\bea
\hspace{-1.5cm}&&
+ \frac{2\,\mgl\,\s2t}{m_t\,(\diff)}\,\left[
\left(4\, m_t^2 + 5\,\msqu + \msqd\right)\,\myltu 
- 2\, \frac{m_t^2\,\msqd}{\msqu}\,\mylg \right]\nn \\
\hspace{-1.5cm}&&\nn\\
\hspace{-1.5cm} &&
- 4\, \frac{\mgl^2 + m_t^2}{\msqu}\mylg 
- 2\,\ln\frac{\msqu}{\mgl^2} + \frac{2}{\den}\left[
2\,\mgl^2\,\left(\mgl^2 + m_t^2 - \msqu \right)
\,\ln\frac{\msqu}{\mgl^2}\nn \right.\nn\\
\hspace{-1.5cm}&&\left.
+ \,2\,m_t^2\,\left(3\,\mgl^2 + 2\,m_t^2 -\msqu 
+ \frac{\mgl^2\,m_t^2-m_t^4}{\msqu}
\right)\,\ln\frac{m_t^2}{\mgl^2}\;\right]
- \frac{4\,\mgl\,m_t\,\s2t}{\msqu\,\den}\,\left[
2\,\msqu\mgl^2 \,\ln\frac{\msqu}{\mgl^2} \right.\nn\\
\hspace{-1.5cm}&&\nn\\
\hspace{-1.5cm} && \left.
-\left((m_t^2-\msqu)^2-\mgl^2\,(m_t^2+\msqu)\right)
\,\ln\frac{m_t^2}{\mgl^2} \right] 
- \frac{8\, \mgl\,m_t}{\s2t\,(\diff)} \, \left[
\myltu - \mylt\myltu \right]\nn\\
\hspace{-1.5cm}&&\nn\\
\hspace{-1.5cm}&& 
- \frac{\mgl\,\s2t}{m_t\,(\diff)}\,\left[
\left( \msqu + \msqd\right)\,\myltuq
+ \left(10\,m_t^2 - 2\,\mgl^2 + \msqu + \msqd\right)\,\mylt\myltu 
\right.\nn\\
\hspace{-1.5cm}&&\nn\\
\hspace{-1.5cm}&& \left.
+\left( 2\,\mgl^2 - 2\,m_t^2 + \msqu + \msqd\right)\,\myltu\mylg
\; \right]
+ \left[\frac{8\,\mgl^2\,m_t^2}{\den}
-\frac{8\,\mgl\,m_t}{\s2t\,(\diff)}\right.
\nn\\
\hspace{-1.5cm}&&\nn\\
\hspace{-1.5cm}&& \left.
+\frac{2\,\s2t\,(4\,m_t^2 \,\mgl^2 - \den)}{\mgl\,m_t\,(\diff)}
+\frac{\s2t\,(\msqu-\mgl^2-m_t^2)^3}{\mgl\,m_t\,\den}\right]\,
\Phi\,(m_t^2,\msqu,\mgl^2) \, , \hspace{1cm}
\label{f2picc}
\eea

\vskip 0.5cm


\bea
\hspace{-1.5cm}f_3(m_t\,,\mgl\,,\msqu\,,\msqd\,,\s2t\,,Q) & = &
- 4\,\frac{\msqd\,(\mgl^2 + m_t^2)}{\msqu\,(\diff)}
+ \frac{4\, \mgl\,m_t\,\s2t}{(\diff)^2} \left(
21\,\msqu - \frac{\msqdq}{\msqu}\;\right) \nn
\eea
\bea
\hspace{-1.5cm}&& +\frac{4}{\diff}\left[ 
\frac{\mgl^2\msqd}{\msqu}\,\mylg
- 2\,(m_t^2 + \mgl^2)\,\myltu \right]
- \frac{24\,\mgl\,m_t\,\s2t\,(3\,\msqu + \msqd)}{(\diff)^2}\,\myltu \nn\\
\hspace{-1.5cm}&&\nn\\
\hspace{-1.5cm}&&
+ \frac{4\,m_t^2}{\msqu\den} \left[
2 \, \mgl^2\,\msqu\,\myltu 
-\mgl^2\,(\mgl^2 -m_t^2 +\msqu)\,\mylg \right. \nn\\
\hspace{-1.5cm}&&\nn\\
\hspace{-1.5cm}&&\left.
- \left((m_t^2 - \msqu)^2-\mgl^2\,(m_t^2+\msqu)\right)\,\mylt 
\,\right]
- \frac{4\,\mgl\,m_t\,\s2t}{\msqu\den} \left[
m_t^2\,(\mgl^2 -m_t^2+\msqu)\,\mylt \right.\nn\\
\hspace{-1.5cm}&&\nn\\
\hspace{-1.5cm}&& \hspace{4cm}\left. 
- \mgl^2\,(\mgl^2 -m_t^2-\msqu)\,\mylg 
+ \msqu\,(\mgl^2 +m_t^2-\msqu)\,\myltu \right]\nn\\
\hspace{-1.5cm}&&\nn\\
\hspace{-1.5cm}&&
+ 2\,\frac{2\,\mgl^2 + 2\,m_t^2-\msqu-\msqd}{\diff}
\,\ln\frac{m_t^2\,\mgl^2}{Q^4}\,\myltu
+ \frac{12\,\mgl\,m_t\,\s2t}{(\diff)^2} \left[
2\,(\mgl^2 - m_t^2)\,\ln\frac{\mgl^2}{m_t^2}\,\myltu\right.\nn\\
\hspace{-1.5cm}&&\nn\\
\hspace{-1.5cm}&&\left.
+(\msqu + \msqd)\,\ln\frac{m_t^2\,\mgl^2}{Q^4}\,\myltu \right]
+ \frac{8\,\mgl\,m_t}{\s2t\,(\diff)^2} \left[
-8\,\msqu + 2\,(3\,\msqu + \msqd)\myltu \right.\nn\\
\hspace{-1.5cm}&&\nn\\
\hspace{-1.5cm}&& \hspace{3.5cm}\left.
-2\,(\mgl^2 - m_t^2)\,\ln\frac{\mgl^2}{m_t^2}\,\myltu
-(\msqu + \msqd)\,\ln\frac{m_t^2\,\mgl^2}{Q^4}\,\myltu \right]\nn\\
\hspace{-1.5cm}&&\nn\\
\hspace{-1.5cm}&&
- \left[
\left(\frac{8}{\s2t} - 12\,\s2t\right)\,
\frac{m_t\,\left( 2\,\den + (\mgl^2+m_t^2-\msqu)\,(\diff)\right)}
{\mgl\,(\diff)^2}
+\frac{4\,\den + 8\,\mgl^2\,m_t^2}{\mgl^2\,(\diff)} \right.\nn\\
\hspace{-1.5cm}&&\nn\\
\hspace{-1.5cm}&&\hspace{1cm}\left.
+\frac{2\,(\mgl^2+m_t^2-\msqu)}{\mgl^2}
-\frac{4\,m_t^2\,(\mgl^2+m_t^2-\msqu - 2\,\mgl\,m_t\,\s2t)}
{\den}\;\right]
\Phi\,(m_t^2,\msqu,\mgl^2) \, ,
\label{f3picc}
\eea

\vskip 0.5cm

\noindent
where $\Delta = \mgl^4 + m_t^4 + \msquq 
- 2\,(\mgl^2\,m_t^2 + \mgl^2\,\msqu + m_t^2\,\msqu)$, 
and the function $\Phi$ is defined as in \cite{davtau}:

\be
\label{defphi}
\Phi\,(x,y,z) = \frac{1}{\lambda} \left(
2\,\ln x_+\,\ln x_- - \ln u \,\ln v -
2\, ( {\rm Li}_2 \,x_+ + {\rm Li}_2 \,x_- ) + \frac{\pi^2}{3} \right) \, ,
\ee
where the auxiliary (complex) variables are:
\be
u = \frac{x}{z}\,,\;\;\;\;\;\;
v = \frac{y}{z}\,,\;\;\;\;\;\;
\lambda = \sqrt{(1-u-v)^2 - 4 \,u\, v}\,,\;\;\;\;\;\;
x_{\pm} = \frac{1}{2}\,(1 \pm (u-v) - \lambda) \, .
\ee
The definition (\ref{defphi}) is valid for the case $x/z < 1$ and 
$y/z < 1$. The other branches of $\Phi$ can be obtained using the symmetry 
properties:
\be
\Phi\,(x,y,z) = \Phi\,(y,x,z)\,,\hspace{1cm}
x\,\Phi\,(x,y,z) = z \, \Phi\,(z,y,x) \, .
\ee
\vskip -1cm
\end{appendletterA}
\begin{appendletterB}
\section*{Appendix B: Shifts of the parameters to the on--shell scheme}
In the OS renormalization scheme, the masses of all particles are 
defined as the poles of the corresponding propagators. As an example, 
for a scalar particle with squared mass $m^2$ the relation between 
the $\overline{\rm DR}$ and  OS definitions of the mass is:  
\be
\delta m^2 = (m^2)^{\overline{\rm DR}} - (m^2)^{\rm OS} = 
{\rm Re}\,\hat{\Pi}\,(m^2) \, ,
\ee
where $\hat{\Pi}\,(m^2)$ is the 
finite part of the self-energy of 
the particle, evaluated at an external momentum equal to the mass itself. 
In the following we list the shifts to the on--shell scheme for the top and
stop masses~\footnote{Similar results have been presented in 
\cite{donini,bagger}. Formulae for the $\overline{\rm DR}-$OS shifts,
specialized to the case $m^2_L = m^2_R$, can also be found in 
\cite{reconciling}. However, we disagree with \cite{reconciling} 
on the contributions to the stop self--energies coming from the 
diagrams that involve the ${\cal O}(\as)$ four--squark vertex.}:
\bea
\frac{\delta \mt}{\mt} & = & \frac{g_s^2}{16 \pi^2} \,C_F\,
\left\{ 3\, \ln \frac{\mt^2}{Q^2} + \delta 
+ \frac{\mgl^2}{\mt^2} 
\left(  \ln \frac{\mgl^2}{Q^2} -1 \right) 
- \frac12 \left[ 
\frac{\msqu}{\mt^2} \left(  \ln \frac{\msqu}{Q^2} -1 \right) 
\right. \right. \nn \\
&&\nn\\
&& \left. \left. - \frac{\mgl^2 +\mt^2-\msqu -2 \,\s2t \,\mgl\, \mt}{\mt^2} 
\,\hat{B}_0(m_t^2,\mgl^2,\msqu) +(1,\s2t) \leftrightarrow (2, -\s2t)
\right] \right\} \, , \label{dmt}\\
&&\nn\\
\frac{\delta\msqu}{\msqu} & = &  \frac{g_s^2}{16 \pi^2} \,C_F\,
 \left\{ 3\, \ln \frac{\msqu}{Q^2} - 7 -
 c^2_{2 \theta} \left( \ln \frac{\msqu}{Q^2} -1 \right) 
-\s2t^2 \frac{\msqd}{\msqu}\left( \ln \frac{\msqd}{Q^2} -1 \right)
 \right. \nn \\
&&\nn\\
&& + 2 \left[  
\frac{\mgl^2}{\msqu} \left(\mylg -1\right)
+ \frac{\mt^2}{\msqu} \left(\mylt -1\right)\right. \nn \\
&&\nn\\
&&\left.\left.\hspace{3.3cm}
+ \frac{\msqu -\mgl^2 -\mt^2 + 2 \,\s2t \,\mgl\, \mt}{\msqu} 
\,\hat{B}_0(\msqu,m_t^2,\mgl^2) \right] \right\} \, , 
\label{dmt1} \\
&&\nn\\
\frac{\delta\msqd}{\msqd} & = & \frac{\delta\msqu}{\msqu} 
\;\; \left[ \left(1,\s2t\right) \leftrightarrow 
\left( 2, -\s2t \right) \right] \, ,
\label{dmt2}
\eea
where the notation $(1,\s2t) \leftrightarrow (2, -\s2t)$ 
in eq.(\ref{dmt}) means a term
that is obtained from the previous ones inside the square bracket
with the exchange $\msqu \leftrightarrow \msqd$ and the replacement
$\s2t \rightarrow -\s2t$. The notation of  
Eq.~(\ref{dmt2}) implies that $\delta\msqd/\msqd$ 
can be obtained from the right hand side of 
eq.~(\ref{dmt1}) with the above substitutions.
The quantity $\delta$ that appears in eq.~(\ref{dmt}) is a constant 
that depends on the regularization. In dimensional regularization 
$\delta = -4$, while in dimensional reduction $\delta = -5$.
In eqs.~(\ref{dmt})--(\ref{dmt2}), $\hat{B}_0$ denotes the finite 
part of the Passarino-Veltman function, i.e.~:
\bea
\hat{B}_0\,(p^2,m_1^2,m_2^2) & = & 
- \int^1_0 dx \ln 
\frac{(1-x)\, m_1^2 + x\,m_2^2 - x\,(1-x)\,p^2 - i\epsilon}{Q^2} \, .
\eea
An explicit expression for $\hat{B}_0$ can be found e.g. in \cite{DS}.

Due to the relative freedom in the choice of the renormalization conditions 
for the stop sector, several definitions are possible for the shift in the 
mixing angle $\thz$ (for a discussion on this point, see \cite{yam} and 
references therein). We choose the following ``symmetrical'' definition: 
\be
\delta\thz =  \frac{1}{2}\, 
\frac{\hat{\Pi}_{12}(\msqu) + \hat{\Pi}_{12}(\msqd)}{\msqu-\msqd} \, ,
\ee
where $\hat{\Pi}_{12}(p^2)$ is the off--diagonal self--energy of the 
stops:
\bea
\hat{\Pi}_{12}(p^2) & = &
\frac{g_s^2}{16 \pi^2} \,C_F\,  \biggr\{ 
4\,m_t\,\mgl\,c_{2\theta}\,\hat{B}_0\,(p^2,m_t,\mgl) \nn \\
&& \hspace{1cm} + c_{2\theta}\,\s2t\,\left[
\msqu \left(  \ln \frac{\msqu}{Q^2} -1 \right)
- \msqd \left(  \ln \frac{\msqd}{Q^2} -1 \right) \right]
\biggr\} \, .
\eea
Finally, taking into account that $\mu$ and $\tan\beta$ do not get any 
${\cal O}\,(\as)$ correction, the shift for the soft term $A_t$ can be 
easily derived from the (field--independent) definition of $\s2t$,
eq.~(\ref{sinmin}):
\be
\label{dat}
\delta A_t = \left( 
\frac{\delta\msqu - \delta\msqd}{\msqu - \msqd} +
\frac{\delta \s2t}{\s2t} - \frac{\delta m_t}{m_t}\right) 
\,(A_t + \mu \cot\beta) \, ,
\ee
where $\delta \s2t/\s2t =  2 \, \cot 2\thz \, \delta\thz$. 
Eq.~(\ref{dat}) can be treated as a definition of $A_t$ in our 
on--shell scheme.
\end{appendletterB}


\setcounter{section}{1}
\begin{appendletterC}
\section*{Appendix C: Two-loop corrections to the CP-odd Higgs mass}

We present here the two--loop ${\cal O}(\at \as)$ corrections to the
mass of the CP--odd Higgs boson, $A$. As apparent from eq.~(\ref{mpij}),
$m_A^2$ depends on the value of the soft supersymmetry--breaking 
parameter $m_3^2$, thus we included it among the MSSM input parameters.
However, a calculation of the loop corrections to the relation between 
$m_A^2$, $m_3^2$ and $\tan \beta$ may be useful for discussing models 
that predict the values of the soft supersymmetry--breaking masses,
and in particular $m_3^2$.

Starting from eqs.~(\ref{mpij}) and (\ref{mpma}), and following the 
same line of reasoning as in Section 3, we find  that:
\be
\label{mam3}
m_A^2 = \frac{1}{\sin\beta\,\cos\beta} \,\left(
- m_3^2 + \frac{h_t^2 \,\mu\,A_t}{\diff}\, F_A \right)\,,
\ee
where $m_A$ is the ${\cal O}(\at\,\as^n)$ loop--corrected $A$ mass 
evaluated in the effective potential approach, i.e. neglecting the 
corrections that depend on the external momenta. The function $F_A$ 
can be decomposed as $F_A = \widetilde{F}_A + \Delta \widetilde{F}_A\,,$ 
where $\Delta \widetilde{F}_A$
contains terms coming from the renormalization of the parameters 
that multiply $F_A$ in eq.~(\ref{mam3}), and $\widetilde{F}_A$ is 
defined as:
\be
\label{fadef}
\widetilde{F}_A =
\frac{\partial V}{\partial \msqd} 
-\frac{\partial V}{\partial \msqu}
+ \frac{4\,\csenza}{\diff}\,\frac{\partial V}{\partial \c2t2}
- \frac{\;\;2\,\mu\,\cot\beta\,(\s2t)^{-2}}{A_t\,(\diff)}
\,\frac{\partial V}{\partial\,\cpp}\;.
\ee
The definitions of the field--dependent parameters $\msqu\,,\;\msqd\,,\;
\c2t2\,$ and $\cpp$ are given in Section 3. The one--loop ${\cal O}(\at)$ 
contribution to $F_A$ is known \cite{erz2}:

\be
\label{fa1l}
F_{A}^{1\,\ell} = \frac{N_c}{16\,\pi^2}\,
\left[\msqu\,\left(1-\myltu\right) - \msqd\,\left(1-\myltd\right)\right]\;.
\ee

\noindent
Assuming a $\overline{\rm DR}$ renormalization for the parameters 
$h_t\,,\;A_t\,,\;\msqu$ and $\msqd$ that enter in eqs. (\ref{mam3}) and
(\ref{fa1l}), the two--loop ${\cal O}(\at\,\as)$ contribution to $F_A$ 
in the $\overline{\rm DR}$ renormalization scheme is given, in units 
of $g_s^2\,C_F\,N_c/(16\,\pi^2)^2$, by:

\bea
\hspace{-1.5cm} \hat{F}_{A}^{2\,\ell} & = &
\frac{16\,\pi^2}{N_c} \,F_{A}^{1\,\ell}\, 
\left[8 - \s2t^2\left(2 -\frac{\msqu+\msqd}{\msqu-\msqd}\,
\ln\frac{\msqu}{\msqd}\right)\;\right]\nn\\
\hspace{-1.5cm}&&\nn\\
\hspace{-1.5cm} &&
+ 2\,\left(\msqu\,\myltuq - \msqd\,\myltdq\right)
+ \frac{2}{\diff}\left(\msqu\,\myltu - \msqd\,\myltd \right)^2\nn\\
\hspace{-1.5cm}&&\nn\\
\hspace{-1.5cm} &&
+ \; f_A(m_t\,,\mgl\,,\msqu\,,\msqd\,,\s2t\,,Q)\; 
- \; f_A(m_t\,,\mgl\,,\msqd\,,\msqu\,,-\s2t\,,Q)\;.
\eea
The function $f_A$ contains contributions coming from the top-stop-gluino
diagrams (fig. \ref{figdiag}d), and its explicit expression in units of
$g_s^2\,C_F\,N_c/(16\,\pi^2)^2$ is:

\bea
f_{A}(m_t\,,\mgl\,,\msqu\,,\msqd\,,\s2t\,,Q) & = &
\frac{ 16\,\msqu\,\mgl\,m_t\,\s2t}{\diff}\nn
- \frac{\pi^2\,\mgl\,(\msqu+\msqd)}{6\,A_t}
- 4 \left(\mgl^2 + m_t^2\right) \myltu
\eea
\vskip -.5cm
\bea
\hspace{-2.5cm} &&
-\frac{2\,\mgl}{A_t}\,
\left[ \msqu\,\left(6-5\,\myltu\right)+\msqd\,\left(1-\myltd\right)\right]
-\frac{4\,\mgl\,m_t\,\s2t\,(3\,\msqu+\msqd)}{\diff}\,\myltu\nn\\
\hspace{-2.5cm}&&\nn\\
\hspace{-2.5cm} &&
-\frac{\mgl}{A_t}\left(\msqu\,\myltuq + \msqd\,\myltdq\right)
+ 2\,(\mgl^2+m_t^2-\msqu)\,\ln\frac{\mgl^2\,m_t^2}{Q^4}\,\myltu\nn\\
\hspace{-2.5cm}&&\nn\\
\hspace{-2.5cm} &&
+ 2\,\msqu\,\left(1+\frac{\mgl}{A_t}\right)\mylg\,\mylt 
- \frac{2\,\mgl}{A_t}\,\left[(\mgl^2-m_t^2)\ln\frac{\mgl^2}{m_t^2}
+\msqu \,\ln\frac{\mgl^2\,m_t^2}{Q^4}\;\right]\,\myltu\nn\\
\hspace{-2.5cm}&&\nn\\
\hspace{-2.5cm} &&
+ \frac{2 \,\mgl\,m_t\,\s2t}{\diff}\left[
2\,(\mgl^2-m_t^2)\ln\frac{\mgl^2}{m_t^2}
+(\msqu+\msqd) \,\ln\frac{\mgl^2\,m_t^2}{Q^4}\;\right]\,\myltu
\nn\\
\hspace{-2.5cm}&&\nn\\
\hspace{-2.5cm} &&
-\left[4\,m_t^2
+ \frac{2\,\den}{\mgl^2}\left(1 + \frac{\mgl}{A_t}\right) 
- \frac{2\,m_t\,\s2t \left(2\,\den + (\mgl^2+m_t^2-\msqu)\,(\diff)\right)}
{\mgl\,(\diff)} \right] \Phi(m_t^2,\msqu,\mgl^2) \, , \nn\\
\eea
where $\den$ and $\Phi(x,y,z)$ are defined at the end of appendix A.

\end{appendletterC}


\end{document}